\documentstyle[aps]{revtex}
\input epsf.sty

\preprint{UCTP-113-98}

\begin{document}
\draft

\title{The effective potential of composite fields  in weakly 
coupled QED in a uniform external magnetic field}

\author{D.-S.~Lee}
\address{Department of Physics, National Dong Hwa University, 
Shoufeng, Hualien 974, Taiwan, ROC}

\author{P.~N.~McGraw, Y.~J.~Ng}
\address{Department of Physics  and Astronomy, University of North 
Carolina, Chapel Hill, North Carolina 27599}

\author{I.~A.~Shovkovy\thanks{
              On leave of absence from Bogolyubov Institute 
              for Theoretical Physics, Kiev 252143, Ukraine}}
\address{Physics Department, University of Cincinnati, 
         Cincinnati, OH 45221-0011, USA}

\date{\today}
\maketitle

\begin{abstract}
The effective potential for the composite fields responsible for 
chiral symmetry breaking in weakly coupled QED in a magnetic 
field is derived.  The global minimum of the effective potential is 
found to acquire a non-vanishing expectation value of the composite 
fields that leads to generating the dynamical fermion mass by an external
magnetic field. The results are compared with those for the 
Nambu-Jona-Lasinio model.  
\end{abstract}

\section{Introduction}

The problem of symmetry breaking and vacuum instability under
the influence of external fields has attracted a lot of attention in
the past few years \cite{JackK,Git}. Recently, some progress was
attained in  understanding the mechanism of the so-called catalysis
of chiral  symmetry breaking
\cite{GMS1,GMS2,Jack,Hong,Vic,Bar,STEZh,Other}  (for some earlier
studies see also \cite{Kle,Klim,Schr}).  This problem provides an
especially interesting example of vacuum engineering:  i.e.,
manipulating external fields to alter the symmetry properties  of
the vacuum.  It is also one of a small number of tractable
non-perturbative problems in (3+1)-dimensional quantum field
theories, and as such may provide insight into the non-perturbative
vacuum structure of other theories such as quantum chromodynamics.

Here we continue the study of this effect in weakly coupled quantum
electrodynamics (QED). In particular, we derive the expression for
the effective potential  of composite fields responsible for chiral
symmetry breaking by  applying the method of \cite{Mir,Shp},
originally introduced in the gauged Nambu-Jona-Lasinio (NJL) model
and QED$_3$. 

In this paper we are interested in constructing the effective 
potential for the composite local fields $\sigma(x)=\langle 0|
\bar{\psi}(x)\psi(x)|0\rangle$ and $\pi(x)=\langle 0|\bar{\psi}(x)
i\gamma_5\psi(x) |0 \rangle$. In general, this problem requires 
consideration of the corresponding generating functional,
\begin{equation}
Z(J_s, J_p)\equiv\exp\left[iW(J_s, J_p)\right]
=\int \prod_{x} \left(d\psi(x) d\bar{\psi}(x) dA_{\mu}(x) \right)
\exp\left[i\int d^4 x\left({\cal L}+J_s(x)\bar{\psi}(x)
\psi(x)+J_p(x)\bar{\psi}(x)i\gamma_5\psi(x)\right)\right].
\label{gen-fun}
\end{equation}
where  ${\cal L}$ is the Lagrangian density of massless QED in an
external  magnetic field. Then the calculation of the effective
action reduces to obtaining  the Legendre transform of the
functional $W(J_s, J_p)$ with respect  to the external sources,
\begin{equation}
\Gamma(\sigma, \pi)=W(J_s, J_p)-\int d^4 x \left[J_s(x)\sigma(x)
+J_p(x)\pi(x)\right].
\label{eff-act}
\end{equation}
Here the sources are functions of fields, which are obtained by
inverting the following expressions:
\begin{equation}
\frac{\delta W}{\delta J_s(x)}=\sigma(x),\qquad 
\frac{\delta W}{\delta J_p(x)}=\pi(x).
\label{del-W}
\end{equation}
The latter, in turn, mean that 
\begin{equation}
\frac{\delta \Gamma}{\delta \sigma(x)}=-J_s(x),\qquad 
\frac{\delta \Gamma}{\delta \pi(x)}=-J_p(x).
\label{del-G}
\end{equation}

\section{The effective potential for composite fields}

Due to the presence of chiral symmetry, the effective potential
is a function of the only invariant, $\rho=\sqrt{\sigma^2+\pi^2}$.
Thus, it is sufficient to keep only one of the sources to be 
non-zero (we choose $J_s\equiv J$, and $J_p=0$). At the end,
the effective potential of one variable $\sigma$ could be 
promoted to the complete expression by the substitution 
$|\sigma |\to\rho$.

Since the constant source (this is just enough for obtaining the 
effective potential) enters the expression for the generating 
functional in Eq.~(\ref{gen-fun}) in exactly the same way as a 
bare mass would ($m_0=-J$), the calculation could be 
considerably simplified \cite{Mir,Shp}.

Indeed, from the definition of the $\sigma=\langle 0|
\bar{\psi}\psi|0\rangle_{J}$ and the first relation 
in Eq.~(\ref{del-W}), we obtain \cite{Mir}
\begin{equation}
w(J)=\int \limits^{\Sigma_0(J)}\langle 0|\bar{\psi}
\psi|0\rangle_{J}\frac{dJ}{d\Sigma_0}d\Sigma_0 ,
\label{w-J}
\end{equation}
where, by definition, $\Sigma_0$ is the value of self-energy
function  at $p=0$, and
\begin{equation}
W(J)\equiv w(J) \int d^4 x .
\label{def-w}
\end{equation}

The chiral condensate is given by the trace of the fermion
propagator. In weakly coupled QED, it is adequate to use the
so-called  lowest Landau level approximation.  In this
approximation, the fermion  propagator (with the momentum dependent
dynamical mass $\Sigma(p_{\parallel})$)  is   given by  
\begin{equation}
S^{(LLL)}(p)=i\exp\left(-p^2_{\perp}l^2\right)
\frac{\not{\!\! p}_{\parallel}+\Sigma(p_{\parallel})}
{p^{2}_{\parallel}-\Sigma^{2}(p_{\parallel})}
\left(1-i\gamma^1\gamma^2\right),
\label{LLL-prop}
\end{equation}
where $l\equiv 1/\sqrt{|eB|}$ is the magnetic length and,
without  loss of generality, we assume that $\mbox{sign}(eB)=1$.
Here we have adopted the same  notations as in Refs.~{\cite{GMS2,Jack}}.
Also, it  is reasonable to use a sharp momentum cut-off at around the
scale  of magnetic field, i.e. $\Lambda\sim\sqrt{|eB|}$. The latter 
means that all the ``bare" quantities in such an approach are defined 
at the characteristic scale of magnetic field. 

The main argument in support of such an approximation is the 
presence of the large Landau gap of order $\sqrt{|eB|}$  in the
energy spectrum for the modes  from the higher Landau levels
\cite{GMS1}, and the absence of any  (or the presence of a small
one after dynamical fermion mass  is generated \cite{GMS2,Jack} ) 
for the modes in the lowest Landau level. Besides, the infrared
region with $ k << \sqrt{eB}$ has been shown to play a dominant 
role in the catalysis of chiral symmetry breaking by an external
magnetic field \cite{GMS1,GMS2,Jack}. 

In the lowest Landau level approximation, from Eq.~(\ref{LLL-prop}) 
we derive
\begin{equation}
\langle 0|\bar{\psi} \psi|0\rangle_{J}=
-\frac{1}{4\pi^2 l^2}
\int \frac{d(k^2)\Sigma(k)}{k^2+\Sigma^2(k)},
\label{def-con}
\end{equation}
where we switched to Euclidean space. We would like to mention here
that the  above expression for the chiral condensate   can also be 
obtained  by making use of  the $E_p $ representation (see Appendix
B in the second paper of Ref.~\cite{Jack}). The Schwinger-Dyson 
equation for the self-energy function, in the same approximation, 
reads \cite{Jack} (compare also with Eq.~(106) in the second paper 
of Ref.~\cite{GMS2})
\begin{equation}
\Sigma(p)=-J+\frac{\alpha}{2\pi^2}
\int \frac{d^2 k \Sigma(k)}{k^2+\Sigma^2(k)}
\int\limits_{0}^{\infty} 
\frac{dq \exp\left(-ql^2/2\right)}{(k-p)^2+q}.
\label{SD1}
\end{equation}
Note that here we added the bare mass term $m_0=-J$, which is
absent in \cite{GMS2,Jack}.

To find a non-perturbative solution to the above equation for the 
self-energy, we use the method of \cite{GMS2} (see Appendix~C of
the second paper). As is common in solving such integral equations,
we approximate it by its linearized version. While looking for a 
solution in the form $\Sigma=\Sigma(k^2)$, we can perform the
angular  integration in Eq.~(\ref{SD1}) exactly. Therefore, we
arrive at
\begin{equation}
\Sigma(p^2)=-J+\frac{\alpha}{2\pi}
\int\limits_{0}^{\Lambda^2} \frac{d(k^2) \Sigma(k^2)}
{k^2+\Sigma_0^2} \int\limits_{0}^{\infty} \frac{dq 
\exp\left(-ql^2/2\right)}{\sqrt{(k^2+p^2+q)^2-4k^2p^2}}.
\label{SD2}
\end{equation}
Finally, after approximating the kernel by its asymptotes 
in two regions $k^2\ll p^2$ and $k^2\gg p^2$ and dropping 
the exponential in the second integral\footnote{We would like 
to thank V.P.~Gusynin for suggesting this approximation.}
(which is justified in our theory with a sharp cut-off at 
the scale of magnetic field), we arrive at the following 
integral equation,
\begin{equation}
M(x)=-j-\frac{\alpha}{2\pi} \ln(x)
\int\limits_{0}^{x} dy \frac{ M(y)}{y+M_0^2}
-\frac{\alpha}{2\pi}
\int\limits_{x}^{1} dy \ln(y) \frac{M(y)}{y+M_0^2},
\label{SD}
\end{equation}
where we use the dimensionless quantities
\begin{equation}
M=\Sigma l, \quad
j=J l, \quad
x=p^2 l^2, \quad
y=k^2 l^2.
\label{dim-less}
\end{equation}

It is straightforward to check that the integral equation
(\ref{SD})  is equivalent to the following differential equation:
\begin{equation}
\frac{d^2 M(x)}{d^2 x}+\frac{1}{x}\frac{d M(x)}{d x}
+\frac{\alpha}{2\pi}\frac{M(x)}{x(x+M_0^2)}=0,
\label{dif-eq}
\end{equation}
with a solution subject to the following (infrared and ultraviolet) 
boundary conditions:
\begin{equation}
\left.\left(xM(x)\right)\right|_{x=0}=0, \qquad
\left.\left(M(x) -x \ln(x)\frac{d M(x)}{d x}\right)
\right|_{x=1}=-j.
\label{bound-c}
\end{equation}

Note also that using the definition of the chiral condensate in 
Eq.~(\ref{def-con}), we arrive at the following expression:
\begin{equation}
\langle 0|\bar{\psi} \psi|0\rangle_{J}=-\frac{1}{4\pi^2 l^3}
\int\limits_{0}^{1} dy \frac{M(y)}{y+M_0^2}
=\frac{1}{2\pi\alpha l^3}
\left.\left(x\frac{d M(x)}{d x}\right)\right|_{x=1},
\label{conden}
\end{equation}
where, to be consistent with the approximations to the
Schwinger-Dyson  equation, we again used the linearized version
instead of the exact  relation.

The solution to Eq.~(\ref{dif-eq}) that is regular at the origin 
(in order to satisfy the infrared boundary condition in 
Eq.~(\ref{bound-c})) is the hypergeometric function \cite{GMS2},
\begin{equation}
M(x)=M_0 F(i\nu,-i\nu;1;-\frac{x}{M_0^2}),\qquad
\nu\equiv\sqrt{\frac{\alpha}{2\pi}}.
\label{solution}
\end{equation}
The ultraviolet boundary condition leads to the following relation
between $M_0$ and the external source:
\begin{equation}
M_0 F(i\nu,-i\nu;1;-\frac{1}{M_0^2})=-j.
\label{UV}
\end{equation}

Now we are in a position to derive the generating functional and 
the effective potential. In weakly coupled QED, the solution for
the self-energy is such  that $M_0\ll 1$ \cite{GMS2,Jack}.
Therefore, using the asymptotic  behavior of the hypergeometric
function \cite{AbSt}, we obtain
\begin{eqnarray}
j&\simeq& -M_0 \sqrt{\frac{\tanh(\pi\nu)}{\pi\nu}} 
\cos\left(\nu\ln(M_0^2)+\Phi(\nu)\right),
\label{asym-j}\\
\langle 0|\bar{\psi}\psi|0\rangle_{J}
&\simeq&\frac{M_0}{4\pi^2\nu l^3}
\sqrt{\frac{\tanh(\pi\nu)}{\pi\nu}} 
\sin\left(\nu\ln(M_0^2)+\Phi(\nu)\right),
\label{asym-cond}
\end{eqnarray}
where we have introduced the following function (which is of order
$\nu^2$  in the coupling)
\begin{equation}
\Phi(\nu)=\mbox{arg}\left(\frac{\Gamma(1-2i\nu)}
{\Gamma^2(1-i\nu)}\right).
\label{def-phi}
\end{equation}
Making use of the expressions in Eqs.~(\ref{asym-j}) and 
(\ref{asym-cond}), we calculate the generating functional,
\begin{eqnarray}
w(j)&=&-\frac{\tanh(\pi\nu)}{8\pi^3\nu^2l^4}\int\limits^{M_0^2(j)}
d \mu \sin\left(\nu\ln(\mu)+\Phi(\nu)\right)
\left[\cos\left(\nu\ln(\mu)+\Phi(\nu)\right)
-2\nu \sin\left(\nu\ln(\mu)+\Phi(\nu)\right)
\right].\nonumber\\
&=&-\frac{M_0^2\tanh(\pi\nu)}{16\pi^3\nu^2l^4}
\left[\sin\left(2\nu\ln(M_0^2)+2\Phi(\nu)\right)-2\nu\right].
\end{eqnarray}
Then the effective potential $V(\rho)$ is given by the 
following parametric representation:
\begin{eqnarray}
V(M_0)&=&j \frac{d w}{d M_0} 
\left(\frac{d j}{d M_0}\right)^{-1}-w(j)
=-\frac{M_0^2\tanh(\pi\nu)}{16\pi^3\nu^2l^4}
\left[\sin\left(2\nu\ln(M_0^2)+2\Phi(\nu)\right)
+2\nu\right], \label{eff-pot}  \\
\rho(M_0)&=&l\frac{d w}{d M_0} 
\left(\frac{d j}{d M_0}\right)^{-1}=\frac{M_0}{4\pi^2\nu l^3}
\sqrt{\frac{\tanh(\pi\nu)}{\pi\nu}} 
\left|\sin\left(\nu\ln(M_0^2)+\Phi(\nu)\right)\right|.
\label{sigma}
\end{eqnarray}
Here, as we argued above, we substituted the chiral invariant 
$\rho$ instead of $|\sigma |$. 

We notice that the effective potential given by
Eqs.~(\ref{eff-pot}),  (\ref{sigma}) is a multivalued function.
This, at first sight, unusual  property of the potential originates
from the presence of an infinite  tower of excited resonances with
the same quantum numbers as those of $\sigma$ and $\pi$ particles.
Among all of these branches, we have to  choose the only one which
corresponds to the stable states. 

\begin{figure}
\epsfbox{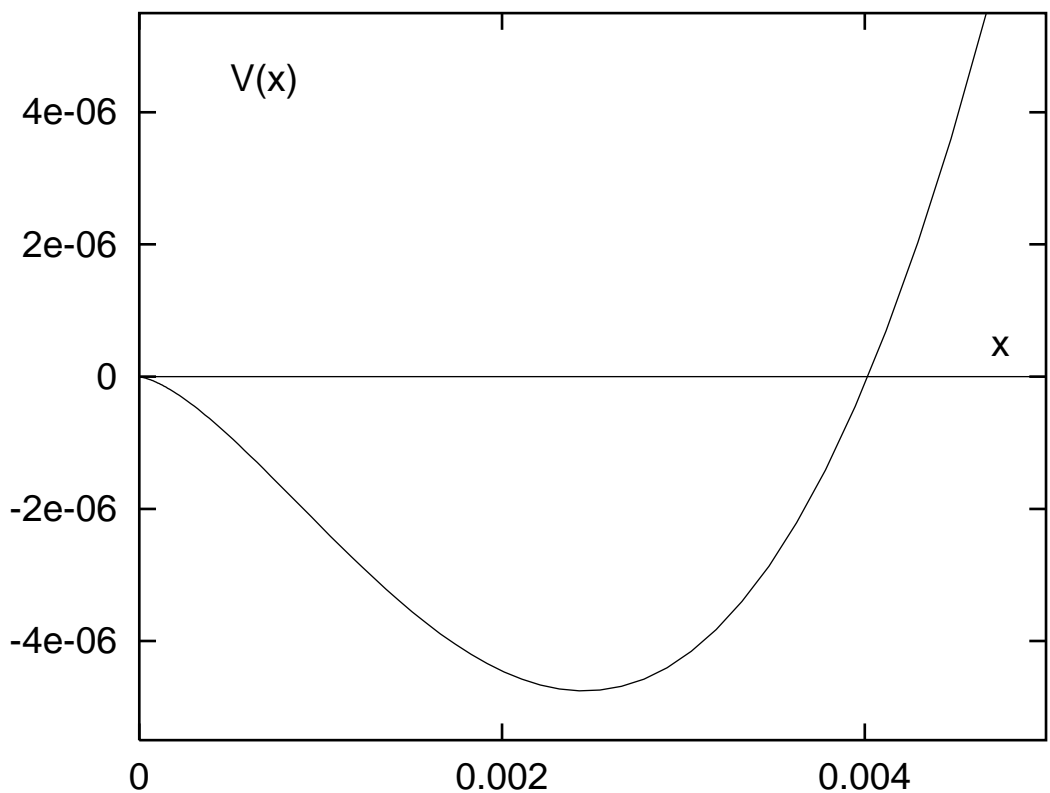}
\caption{The effective potential at $\nu=0.2$ in the vicinity 
of the $n=0$ minimum. The units are chosen so that $l=1$. The $x$ axis
represents the chiral invariant $\rho$.}  
\label{Figure1}
\end{figure}

The gap equation, derived from this effective potential, is
equivalent  to the equation $j=0$. The latter leads to the infinite
set of solutions
\begin{eqnarray}
M^{(n)}_0&=&\exp\left(-\frac{\pi}{4\nu}(2n+1)\right), 
\qquad n=0,1,2,\dots, \label{m_n} \\
\rho^{(n)}&=&\frac{1}{4\pi^2\nu l^3}
\sqrt{\frac{\tanh(\pi\nu)}{\pi\nu}} 
\exp\left(-\frac{\pi}{4\nu}(2n+1)\right).
\label{sol-gap}
\end{eqnarray}
Note that the solutions with negative values of $n$ are all
spurious,  since they correspond to values of the dynamical fermion
mass larger than the cut-off (we remind that $\Lambda=1/l$ here). 
All the  solutions in Eq.~(\ref{sol-gap}) are local minima of the
effective potential. Indeed,
\begin{equation}
\left.\frac{d^2V(\rho)}{d\rho^2}\right|_{\rho^{(n)}}=\frac{1}{l}
\left.\frac{d j}{d M_0}\left(\frac{d \rho}{d M_0}\right)^{-1}
\right|_{\rho^{(n)}}=8 \pi^2 \nu^2 l^2 >0.
\label{sec-der}
\end{equation}
However, the only solution that represents the stable composite 
$\sigma$ and $\pi$ particles is that with $n=0$ \cite{GMS2}. We
could also check that the value of the effective potential at
$\rho^{(0)}$ is lower than any of the values at the other minima,
as it should be. In figure \ref{Figure1} the potential is plotted
in the physically interesting region  in the vicinity of
$\rho^{(0)}$.  In this region, the potential as a function of
$\rho$ is approximately given by the following expression:
\begin{equation}
V(\rho)\simeq-2\pi^2 \nu^2 l^2 \rho^2
\left[1-\ln\left(\frac{\rho}{\rho^{(0)}}\right)^{2}\right].
\label{eff-app}
\end{equation}
Clearly, the dynamical fermion mass $m_{dyn}$ that corresponds to
the global minimum  $\rho^{(0)}$ is
\begin{equation}
m_{dyn}=M_0^{(0)}/l \simeq \sqrt{|eB|}
\exp\left(-\sqrt{\frac{\pi}{\alpha}}\right), 
\end{equation}
which is consistent with the result obtained in
Refs.~{\cite{GMS2,Jack}}.

\section{Conclusion}

In this paper we derived the effective potential of the composite
fields  responsible for chiral symmetry breaking in weakly coupled
QED  in a constant magnetic field. The global minimum of the
effective potential is obtained and is found to  acquire a
non-vanishing expectation value of the composite fields that leads
to generating the dynamical fermion mass by an external magnetic
field. Comparing the results in QED with those in the NJL model
\cite{GMS1}, we notice a qualitative difference in their effective
potentials. While in the NJL model the potential is a single valued
function, in QED it is multivalued. The latter property reflects
the presence of a tower of excited states  with the same quantum
numbers as those of the Nambu-Goldstone bosons.  From the physical
point of view, this difference appears due to  the long-range
interaction in QED in contrast to the short-range  interaction in
NJL model. 

To calculate the full low-energy effective action, we also need the
kinetic  terms for these fields. Applying the general method of
Refs.~\cite{GM,GMK}, the  kinetic part of the effective Lagrangian
density can be expressed in terms of the  propagators of composite
scalar and pseudoscalar fields, $\Delta_s(q)$  and  $\Delta_p(q)$:
\begin{equation}
{\cal L}_{kin}=-\frac{1}{4}
\left.\frac{\partial^2\left(\Delta^{-1}_s(q)\right)}
{\partial q_{\mu}\partial q_{\lambda}}\right|_{q=0}
\partial_{\mu}\tilde{\sigma}(x)
\partial_{\lambda}\tilde{\sigma}(x) -\frac{1}{4}
\left.\frac{\partial^2\left(\Delta^{-1}_p(q)\right)}
{\partial q_{\mu}\partial q_{\lambda}}\right|_{q=0}
\partial_{\mu}\tilde{\pi}(x)\partial_{\lambda}\tilde{\pi}(x).
\label{kin}
\end{equation}
The propagators, entering the right hand side of this formula,
depend on the non-perturbative fermion propagator (the solution
to  the Schwinger-Dyson equation) and the scalar and pseudoscalar
vertex  functions \cite{GM,GMK}.  The calculation of these vertex
functions requires the solution of an additional non-perturbative
self-consistency equation analogous to the Schwinger-Dyson
equation,  which is beyond the scope of the present paper.  We
expect that the forms of the kinetic terms and the corresponding
dispersion relations should be qualitatively similar to those
derived in the case of the NJL model \cite{GMS1}.  Important points
are: 1) there is no mass gap in the pseudoscalar dispersion
relation,  since the pseudoscalar particle is a Goldstone boson. 
2) For the scalar particle there is a mass gap, and we expect that,
as in the NJL model,  the coefficients of the kinetic terms will be
different for the directions parallel and perpendicular to the
applied magnetic field.  The dependence of the dispersion law on
the transverse components is expected to be strongly suppressed, 
but it is nonetheless extremely important, playing the key role in
preventing the chiral symmetry breaking from being washed out. 

\begin{acknowledgments}

We would like to thank V.P.~Gusynin for many interesting
discussions  and suggesting a simple approximation for the kernel
of the integral  equation (\ref{SD2}). I.A.S. acknowledges useful
discussions with  L.C.R.~Wijewardhana. D.S.L. would like to thank
the members of the  department of physics and astronomy of the
University of Pittsburgh and  the department of physics and
astronomy of the University of North Carolina,  Chapel Hill for
their hospitality. The work of I.A.S. was supported by U.S. 
Department of Energy  Grant \#DE-FG02-84ER40153.  The work of
D.S.L. was  supported by the Republic Of China National Science
Council through grant  \#NSC 88-2112-M-259-001.  The work of P.N.M.
and Y.J.Ng was supported by the U.S. Department of Energy through
grant \#DE-FG05-85ER-40219 Task A.

\end{acknowledgments}

\begin{references}

\bibitem{JackK} Y.~J.~Ng and Y.~Kikuchi, in {\sl Vacuum Structure 
in Intense Fields}, ed. by H.~M.~Fried and B.~M\"{u}ller, (Plenum,
N.Y., 1991).

\bibitem{Git} D.~M.~Gitman, E.~S.~Fradkin and 
Sh.~M.~Shvartsman, in {\sl Quantum Electrodynamics with 
Unstable Vacuum}, ed. by V.~L.~Ginzburg (Nova Science, Commack, 
N.Y., 1995).

\bibitem{GMS1} V.~P.~Gusynin, V.~A.~Miransky, and I.~A.~Shovkovy,
\prl {\bf 73}, 3499 (1994); \prd {\bf 52}, 4718 (1995);
\pl B{\bf 349}, 477 (1995).

\bibitem{GMS2} V.~P.~Gusynin, V.~A.~Miransky, and I.~A.~Shovkovy, 
\prd {\bf 52}, 4747 (1995); Nucl.\ Phys.\ B{\bf 462}, 249 (1996).

\bibitem{Jack} C.~N.~Leung, Y.~J.~Ng, and A.~W.~Ackley, \prd
{\bf 54}, 4181 (1996); D.-S.~Lee, C.~N.~Leung, and Y.~J.~Ng,
\prd {\bf 55}, 6504 (1997); \prd {\bf 57}, 5224 (1998).

\bibitem{Hong} D.~K.~Hong, Y.~Kim, and S.-J.~Sin, \prd {\bf 54}, 
7879 (1996); D.~K.~Hong, \prd {\bf 57}, 3759 (1998).

\bibitem{Vic} V.~Elias, D.~G.~C.~McKeon, V.~A.~Miransky, and
I.~A.~Shovkovy, \prd {\bf 54}, 7884 (1996); V.~A.~Miransky, 
hep-th/9805159.

\bibitem{Bar} D.~M.~Gitman, S.~D.~Odintsov and Yu.~I.~Shilnov,
\prd {\bf 54}, 2968 (1996); E.~Elizalde, Yu.~I.~Shil'nov, V.~V.~Chitov,
Class.\ Quant.\ Grav.\ {\bf 15}, 735 (1998).

\bibitem{STEZh} I.~A.~Shovkovy and V.~M.~Turkowski,
Phys.\ Lett.\ B {\bf 367}, 213 (1996); 
D.~Ebert and V.~Ch.~Zhukovsky, 
Mod.\ Phys.\ Lett.\ A {\bf 12}, 2567 (1997).

\bibitem{Other} I.~A.~Shushpanov and  A.~V.~Smilga,
\pl B {\bf 402}, 351 (1997); 
E.~J.~Ferrer and V.~de~la~Incera,
\prd {\bf 58}, 065008 (1998); 
S.~Kanemura,  H.-T.~Sato and H.~Tochimura, 
Nucl.\ Phys.\ B {\bf 517}, 567 (1998);
 A.~Yu.~Babansky, E.~V.~Gorbar and G.~V.~Shchepanyuk,
\pl B {\bf 419}, 272 (1998); 
T.~Itoh and H.~Kato, \prl {\bf 81}, 30 (1998);
K.~Farakos, G.~Koutsoumbas, and N.E. Mavromatos, 
\pl B {\bf 431}, 147 (1998).

\bibitem{Kle} S.~P.~Klevansky and R.~H.~Lemmer, 
\prd {\bf 39}, 3478 (1989).

\bibitem{Klim} K.~G.~Klimenko, Theor.\ Math.\ Phys. {\bf 89}, 
1161 (1992); Z.\ Phys. C{\bf 54}, 323 (1992);  K.~G.~Klimenko, 
B.~V.~Magnitskii and A.~S.~Vshivtsev, 
Nuovo Cim.\ A {\bf 107}, 439 (1994). 
 
\bibitem{Schr} S.~Schramm, B.~M\"{u}ller, and A.~J.~Schramm,
Mod.\ Phys.\ Lett. A{\bf 7}, 973 (1992).

\bibitem{Mir}  V.~A.~Miransky, Int.\ J.\ Mod.\ Phys. A{\bf 8},
135 (1993).

\bibitem{Shp} V.P.~Gusynin, V.A.~Miransky and A.V. Shpagin, 
\prd {\bf 58}, 085023 (1998).

\bibitem{AbSt} {\sl Handbook of Mathematical Functions}, ed.
by M.~Abramowitz and I.~A.~Stegun, (1968).

\bibitem{GM} V.P.~Gusynin and V.A.~Miransky, 
Mod.\ Phys.\ Lett.\ A{\bf 6}, 2443 (1991); 
Sov.\ Phys.\ JETP {\bf 74}, 216 (1992). 

\bibitem{GMK} V.P.~Gusynin, V.A.~Kushnir and V.A.~Miransky,
\prd {\bf 39}, 2355 (1989); Erratum-ibid. \prd {\bf 41}, 3279 
(1990). 

\end{references}
\end{document}